\documentclass[prd,showpacs,showkeys,preprintnumbers,amsmath,amssymb,superscriptaddress,floatfix,nofootinbib]{revtex4}
\usepackage{amssymb}
\usepackage{amsmath}
\usepackage{amsfonts}
\usepackage{epsfig}
\usepackage{graphicx}
\usepackage{tabularx}
\usepackage{color}

\newcommand{\seq}{\begin{subequations}}
\newcommand{\sen}{\end{subequations}}
\newcommand{\eq}{\begin{eqnarray}}
\newcommand{\en}{\end{eqnarray}}
\newcommand{\ra}{\rangle}
\newcommand{\la}{\langle}

\newcommand{\bfb}{{\bf b}_{\perp}}

\begin{document}

\title{Tetraquarks in holographic QCD} 

\author{Thomas Gutsche}
\affiliation{Institut f\"ur Theoretische Physik,
Universit\"at T\"ubingen, \\
Kepler Center for Astro and Particle Physics,
Auf der Morgenstelle 14, D-72076 T\"ubingen, Germany}
\author{Valery E. Lyubovitskij}
\affiliation{Institut f\"ur Theoretische Physik,
Universit\"at T\"ubingen, \\
Kepler Center for Astro and Particle Physics,
Auf der Morgenstelle 14, D-72076 T\"ubingen, Germany}
\affiliation{Departamento de F\'\i sica y Centro Cient\'\i fico
Tecnol\'ogico de Valpara\'\i so-CCTVal, Universidad T\'ecnica
Federico Santa Mar\'\i a, Casilla 110-V, Valpara\'\i so, Chile}
\affiliation{Department of Physics, Tomsk State University,
634050 Tomsk, Russia}
\affiliation{Laboratory of Particle Physics,
Tomsk Polytechnic University, 634050 Tomsk, Russia}
\author{Ivan Schmidt}
\affiliation{Departamento de F\'\i sica y Centro Cient\'\i fico
Tecnol\'ogico de Valpara\'\i so-CCTVal, Universidad T\'ecnica
Federico Santa Mar\'\i a, Casilla 110-V, Valpara\'\i so, Chile}

\date{\today}

\begin{abstract}

Using a soft-wall AdS/QCD approach we derive the Schr\"odinger-type 
equation of motion for the tetraquark wave function, which is dual 
to the dimension-4 AdS bulk profile. 
The latter coincides with the number of constituents in the 
leading Fock state of the tetraquark. The obtained equation of 
motion is solved analytically, providing predictions for both the
tetraquark wave function and its mass.  
A low mass limit for possible tetraquark states is given by
$M \ge 2 \kappa = 1$ GeV, where $\kappa = 0.5$ GeV is the typical value 
of the scale parameter in soft-wall AdS/QCD. We confirm results 
of the COMPASS Collaboration recently reported on the discovery of 
the $a_1(1414)$ state, interpreted as a tetraquark state composed of 
light quarks and having $J^{PC} = 1^{++}$. Our prediction for the mass 
of this state, $M_{a_1} = \sqrt{2}$ GeV $\simeq 1.414$ GeV, is in good 
agreement with the COMPASS result $M_{a_1} = 1.414^{+0.015}_{-0.013}$ GeV.  
Next we included finite quark mass effects, which are essential for the 
tetraquark states involving heavy quarks. 

\end{abstract}



\date{\today}

\maketitle

\section{Introduction}

This work is addressed to the problem of constructing 
hadronic wave functions of tetraquark states using the AdS 
and light-front QCD correspondence~\cite{Brodsky:2007hb}.  
The idea for such a correspondence works very well for 
conventional hadrons --- mesons and baryons~\cite{Brodsky:2007hb,%
Brodsky:2008pf,Brodsky:2011xx,Branz:2010ub,Gutsche:2012ez,%
Gutsche:2014oua,Gutsche:2013zia}.   
In particular, from the matching of matrix elements for physical 
processes one can relate the bulk profile of the AdS field in 
a holographic dimension to the transverse part of the hadronic light-front 
wave function (LFWF) for the case of massless quarks~\cite{Brodsky:2007hb}.    
The LFWF was generalized in Ref.~\cite{Brodsky:2008pg}  
for the case of a two-parton state by the explicit inclusion of 
the constituent quark masses in the LF kinetic energy 
$\sum_i ({\bf k}_{\perp i}^2 + m_i^2)/x_i$ while the introducing of 
quark masses for tetraquark states was done in Ref.~\cite{Brodsky:2016yod}. 
In the LFWF the inclusion of quark masses corresponds to the introduction 
of the longitudinal wave function (WF),  which was done in this particular 
case through the so-called Brodsky-Huang-Lepage (BHL) 
or {\it Gaussian ansatz}~\cite{Brodsky:1982nx,Huang:1994dy}.  
In Refs.~\cite{Branz:2010ub,Gutsche:2012ez,Gutsche:2014oua} 
we studied the problem of the longitudinal part of the LFWF  
following the ideas of Ref.~\cite{Brodsky:2008pg}.     
In particular, in Ref.~\cite{Gutsche:2012ez} we derived the longitudinal part of 
the LFWF, using constraints of chiral symmetry in the sector of light quarks, 
and heavy quark effective theory in the sector of heavy quarks.  
The idea that explicit breaking of chiral symmetry is a property of 
the longitudinal part of the hadronic LFWF and that it can be induced via 
the current quark mass dependence of 
the longitudinal LWFW, had been proposed before in large $N_c$ 
two-dimensional QCD~\cite{'tHooft:1974hx}. This mechanism was later used
in the context of the two-dimensional massive Schwinger 
model~\cite{Bergknoff:1976xr,Ma:1987wi,Mo:1992sv} and was reexamined in 
Refs.~\cite{Chabysheva:2012fe,Forshaw:2012im}. 

In the present paper we extend the soft-wall AdS/QCD model~\cite{SW} 
to the description of tetraquarks --- compact exotic states composed 
of two quarks and two antiquarks  
(for reviews see e.g. Ref.~\cite{Nielsen:2009uh}-\cite{Ali:2017jda}). 
In Refs.~\cite{Ebert:2005nc,Dubnicka:2011mm,Gutsche:2016cml} tetraquarks 
have been analyzed in the context of relativistic quark models. 
Discussions on tetraquark production can be found, e.g. in 
Refs.~\cite{Brodsky:2014xia}-\cite{Brodsky:2016uln}. 
The formalism for tetraquark production 
consistent with quark counting rules has been developed 
in Refs.~\cite{Brodsky:2014xia,Voloshin:2016phx,Brodsky:2016uln}. 
In Ref.~\cite{Chow:1994hg} tetraquarks were studied in the
large $N_c$ limit and it was correctly stressed that 
in the context of $SU(N_c)$ color symmetry the quark $q^a$ and 
antiquark $\bar q^a$ fields can be replaced by $N_c-1$ 
antiquark and quarks, respectively, because of similar transformations  
concerning the $SU(N_c)$ color group. It means that baryons and multiquark 
states emerge from quark-antiquark mesons under the replacements 
$\bar q^{a_1} \to \epsilon^{a_1 \ldots a_n} q^{a_2} \ldots q^{a_n}$ and  
$q^{a_1} \to \epsilon^{a_1 \ldots a_n} \bar q^{a_2} \ldots \bar q^{a_n}$ 
as 
\eq 
& &{\rm Mesons} \to 
N_c \, {\rm Baryons}: \quad  
\bar q^{a} q^{a} \to 
\epsilon^{a b_2 \ldots b_n} q^{a} q^{b_2 } \ldots q^{b_n}\,, 
\nonumber\\
& &{\rm Mesons} \to 
2(N_c-1) \, {\rm Multiquarks}: \quad  
\bar q^{a} q^{a} \to 
\epsilon^{a a_2 \ldots a_n} \epsilon^{a b_2 \ldots b_n} \, 
q^{a_2} \ldots q^{a_n} \, \bar q^{b_2} \ldots \bar q^{b_n} \,. 
\en 
In the case of QCD we have $N_c = 3$ and arrive at a  picture where 
mesons, baryons and tetraquarks appear as fundamental color singlet states. 
This is consistent with arguments of superconformal symmetry where 
mesons, baryons with $L=0$ and $L=1$, tetraquarks are classified 
as members of a superquadruplet~\cite{Brodsky:2016yod}. 

First applications of AdS/QCD to the description of tetraquarks 
have been performed in Refs.~\cite{Forkel:2010gu,Brodsky:2016yod,Guo:2016uaf}. 
In particular, in Ref.~\cite{Forkel:2010gu} the effective 
action for light scalar tetraquarks was derived. 
Based on this action the equation of motion for the 
wave functions and mass spectrum $M^2$ of scalar tetraquarks are derived. 
The spectrum results in  
\eq\label{EOM_Forkel} 
M^2 = 4 \kappa^2 \Big(n + 3\Big)\,, 
\en  
where $\kappa \sim 500$ MeV is the dilaton scale parameter 
in soft-wall AdS/QCD and $n$ is the radial quantum number. 
From Eq.~(\ref{EOM_Forkel}) it follows that 
the lower bound for the ground-state mass of the 
tetraquark is set by $2 \kappa \sqrt{3}$. 
In Ref.~\cite{Brodsky:2016yod} the tetraquark state was introduced 
as a partner of the supersymmetric quadruplet consisting of two baryon 
states with positive and negative chirality, meson state and tetraquark, 
which it is consistent with the ideas of Ref.~\cite{Chow:1994hg}. 
In this vein, it was shown in Ref.~\cite{Brodsky:2016yod} that 
the lowest-lying light-quark tetraquark $f_0(980)$, with 
isospin $I=0$ and spin-parity $J^{PC}=0^{++}$, is a partner of 
the $b_1(1235)$ with $J^{PC}=0^{+-}$ and the nucleon, with  
$J^P=\frac{1}{2}^+$,  
while possible more heavier tetraquarks, the axial state $a_1(1260)$ 
with $J^{PC}=1^{++}$, could be a partner of the
$\Delta(1230)$ with $J^P=\frac{3}{2}^+$ and the $a_2(1320)$ 
with $J^{PC}=2^{++}$. The mass formula for tetraquarks with 
quantum numbers $n, L, S$ (radial quantum number, orbital angular momentum 
and internal spin) derived in Ref.~\cite{Brodsky:2016yod}, 
in the limit of massless quarks reads 
\eq 
M^2 = 4 \kappa^2 \Big(n + L + 1 + \frac{S}{2}\Big) 
\en 
The formalism for the study of tetraquarks proposed in Ref.~\cite{Guo:2016uaf} 
contains a basic error, because of the use of the conformal dimension 
for the two-partonic states.  This gives misleading results for the twist-scaling 
for small values of the holographic coordinate of the tetraquark wave function 
and results in a underestimate of the tetraquark mass spectrum. 
E.g. the lower bound for the tetraquark mass is incorrect. 
In our consideration we derive the action for the tetraquarks 
with adjustable quantum numbers of $n, J, L$. 
Then we derive the equation of motion for the tetraquark 
wave function and the resulting mass spectrum $M^2$ 
\eq 
M^2 = 4 \kappa^2 \Big(n + \frac{L+J}{2} + 1\Big) \,. 
\en  
Our equation of motion and solution for the tetraquark mass spectrum
is different from the ones derived in Ref.~\cite{Forkel:2010gu}, because of 
the different choice for the dimension of the AdS fields dual to tetraquarks. 
In our case the conformal dimension is $\Delta = \tau = N + L$, 
where $\tau_{L=0} = N = 4$ 
is the leading twist-dimension of the tetraquark Fock state 
(corresponding to the number of partons in the leading Fock 
state) and $L$ is the maximal magnitude of the orbital angular momentum 
in the four-quark configuration. Such choice for the conformal 
dimension guarantees the correct power scaling of hadronic 
form factors at large values of the Euclidean transfer momentum squared $Q^2$.  
In Ref.~\cite{Forkel:2010gu}, the conformal dimension was chosen as 
$\Delta = 6$ for the light scalar tetraquark, which differs from 
our choice $\Delta = 4, 5$ at $N = 4$ and $L=0, 1$. 
Moreover, our mass formula is different from the result of 
Ref.~\cite{Brodsky:2016yod} due to a correction which takes 
into account the spontaneous breaking of superconformal symmetry with 
\eq 
\Delta M^2 = 2 \kappa^2 \, (J-L-S) \,. 
\en 
Next we include the finite masses of the constituents 
which form the tetraquark states. Note that the first inclusion of 
finite quark mass corrections to the tetraquark spectroscopy has been 
done in Ref.~\cite{Brodsky:2016yod}. 
In our case we proceed in analogy with 
quark-antiquark mesons (see details in Ref.~\cite{Gutsche:2012ez}), 
where we derived full wave functions containing 
transverse wave functions matched from AdS/QCD 
and longitudinal wave functions encoding the mass effects of tetraquark 
constituents. In particular, we consider two possibilities: 
1) tetraquarks are bound states of two quarks and two antiquarks;  
3) tetraquarks are bound states of two mesons (hadronic molecules). 
 
The paper is organized as follows.
In Sec.~II we present our formalism and at the end derive
the master equation of motion for the full wave function of the tetraquark, 
providing solutions for the mass spectrum including diquark mass effects. 
In Sec.~III we present numerical results and give our final conclusions. 

\section{Approach}

The starting point for a discussion of tetraquarks in 
soft-wall AdS/QCD is the action for spin-$J$ boson fields 
$\Phi_J = \Phi_{M_1 \cdots M_J}(x,z)$, 
with a negative dilaton in the exponential prefactor, 
proposed in Refs.~\cite{Branz:2010ub,SW6} 
\eq 
S_J = \int d^d x dz \sqrt{g} e^{-\varphi(z)}
\Big[ \partial_M\Phi_J \, \partial^M\Phi^J
- (\mu_J^2 + \hat{V}_J(z)) \Phi_J \Phi^J  \Big]\,. 
\en  
This expression is fully equivalent to the action with a positive 
dilaton~\cite{Brodsky:2014yha} after appropriate dilaton-dependent  
redefinition of the spin-$J$ boson fields.                
                                                                   
The AdS metric is specified as 
$ds^2 = e^{2A(z)} (dx_\mu dx^\mu - dz^2)$, 
$g = e^{5A(z)}$, $A(z)=\log(R/z)$ and $R$ is the AdS radius.  
Here $\Phi_J$ is the symmetric, traceless tensor 
classified by the representation $D(E_0,J/2,J/2)$ with 
energy $E_0 = \Delta = \tau$. $E_0$ is related to the bulk 
mass $\mu_J$ as $\mu_J^2 R^2 = (E_0 - J) (E_0 - 4 + J)$  
where $\tau = N + L$ is the twist-dimension and  
$N$ is number of partons; 
$\hat{V}_J(z)$ is the effective dilaton potential, which has an analytical 
expression in terms of the field $\varphi(z)$ and the "metric" field $A(z)$, 
without referring to any specific form of their $z$ profiles:  
\eq 
\hat{V}_J(z) = e^{-2A(z)} \hat{U}_J(z)\,, \quad 
\hat{U}_J(z) = \varphi^{\prime\prime}(z)
+ (d-1-2J) \, \varphi^\prime(z) \, A^\prime(z)  \,.
\en 
In order to study the bound-state problem it is convenient to 
make a dilaton-dependent redefinition of the bulk field, 
$\Phi_J \to \Phi_J e^{\varphi(z)/2}$. 
In the case of the action with positive dilaton profile the 
dilaton-dependent redefinition of the bulk field is 
$\Phi_J \to \Phi_J e^{-\varphi(z)/2}$. 
After the redefinitions $\Phi_J \to \Phi_J e^{\pm\varphi(z)/2}$ 
both versions of the soft-wall model reduce to the same no-wall action 
\eq\label{action_S}
S_J = \int d^d x dz \sqrt{g}
\Big[ \partial_M\Phi_J \, \partial^M\Phi^J  
- (\mu_J^2 + V_J(z)) \Phi_J \Phi^J \Big]\,, 
\en 
where 
\eq 
V_J(z) = e^{-2A(z)} U_J(z)\,, \quad 
U_J(z) = \frac{1}{2} \Big( \varphi^{\prime\prime}(z)
+ \frac{(\varphi^{\prime}(z))^2}{2}+ (d-1-2J) \, \varphi^\prime(z) \, 
A^\prime(z) \Big) \,.
\en 
At $d=4$, $A(z) = \log(R/z)$ and $\varphi(z) = \kappa^2 z^2$ one gets 
\eq 
U_J(z) = \kappa^4 z^2 + 2 \kappa^2 (J-1) \,. 
\en 
The potential $U_J(z)$ is the confinement potential which 
breaks both conformal and chiral invariance spontaneously. 

Next, using the Kaluza-Klein decomposition 
\eq 
\Phi_J(x,z) &=& 
\sum\limits_n \phi_{nJ}(x) \Phi_{n\tau}(z)\,, \nonumber\\  
\Phi_{n\tau}(z) &=& e^{-A(z)\,(d-1)/2} \, \varphi_{n\tau}(z)\,, 
\en 
we derive the Schr\"odinger-type equation for the bulk profiles 
$\varphi_{n\tau}(z)$ 
\eq 
\biggl[ - \frac{d^2}{dz^2} + U_J(z) + W_{\tau}(z) \biggr] \varphi_{n\tau}(z) = 
M^2_{nJ\tau}(z) \varphi_{n\tau}(z) \,,
\en   
where 
\eq 
W_{\tau}(z) = \frac{4 \Delta (\Delta - d) + d^2 - 1}{4z^2} 
= \frac{4 \tau (\tau - d) + d^2 - 1}{4z^2}  
\en 
is the centrifugal potential.  
For $d=4$ and in the case of quark-antiquark mesons ($\tau = 2+L$), 
tetraquarks ($\tau = 4+L$) and 
six quarks ($\tau = 6+L$), this potential reads 
\eq 
& &{\rm Mesons} \quad W_{2+L}(z) = \frac{4L^2-1}{4z^2}\,, \nonumber\\
& &{\rm Tetraquarks} \quad W_{4+L}(z) = \frac{4(L+2)^2-1}{4z^2}\,, \\
& &{\rm Sixquarks} \quad W_{6+L}(z) = \frac{4(L+4)^2-1}{4z^2}\,, \nonumber 
\en   
The hadronic wave functions are identified with the profiles of 
the AdS modes $\varphi_{n\tau}(z)$ in the $z$ direction: 
\eq
\hspace*{-.2cm}
\varphi_{n\tau}(z) = \sqrt{\frac{2 \Gamma(n+1)}{\Gamma(n + \tau - 1)}} 
\kappa^{\tau-1} z^{\tau-3/2} e^{-\kappa^2z^2/2} L_n^{\tau-2}(\kappa^2z^2)\,.  
\en 
They posses the correct behavior in both the ultraviolet and infrared limits, 
with  
\eq
\Phi_{n\tau}(z) \sim z^{3/2} \varphi_{n\tau}(z) 
\to z^{\tau}  \ {\rm at \ small} \ z, \ 
\Phi_{n\tau}(z) \to 0 \  {\rm at \ large} \ z \,
\en 
and are normalized according to the condition 
\eq 
\int\limits_0^\infty dz \varphi_{n\tau}^2(z) = 1 \,. 
\en 
The mass spectrum of multiquark meson states is given by 
\eq 
M^2_{nJ\tau} = 4 \kappa^2 \, \biggl[ n + \frac{\tau + J - 2}{2} \biggr]
= 4 \kappa^2 \, \biggl[ n + \frac{L + J}{2} + 1 \biggr] \,. 
\en 
For $z \to 0$ the scaling of the bulk profile is identified 
with the scaling of the corresponding mesonic interpolating operator 
$\tau$. As we mentioned in the Introduction, $\tau$ 
depends on $L$ (instead of $J$ as in conformal field 
theory), because we are modeling QCD and 
therefore should reproduce the scaling of hadronic form factors. 
As we stressed before, the dependence on $L$ reflects the  a spontaneous 
breaking of chiral invariance, which is expected, 
since after the introduction of the dilaton field we break the conformal 
or gauge invariance acting in AdS space. As we noted 
before, the chiral group is isomorphic to a subgroup of $SO(4,2)$. 

The next step is to include the longitudinal part of the LFWF  
following the approach used in our paper on mesons~\cite{Gutsche:2012ez}. 
In the case of mesons the first step in this 
direction was taken in Refs.~\cite{Brodsky:2008pg} for mesons and 
in Ref.~\cite{Brodsky:2016yod} for tetraquarks. These authors
proposed a factorized form for the mesonic two-parton wave function  
as a product of transverse $\phi_{nL}(\zeta)$, 
longitudinal $f(x,m_1,m_2)$ and angular $e^{im\phi}$
modes. In Ref.~\cite{Branz:2010ub} we presented 
a more convenient form, factorizing the additional factor $\sqrt{x (1-x)}$,
which is the Jacobian of the coordinate transformation $\zeta \to |\bfb|$, with:   
\eq
\psi_{q_1\bar q_2}(x,\zeta,m_1,m_2) =
\frac{\phi_{nL}(\zeta)}{\sqrt{2\pi\zeta}} \, 
f(x,m_1,m_2) \, e^{im\phi} \, \sqrt{x(1-x)} \,. 
\en 
In Ref.~\cite{Branz:2010ub} we used a {\it Gaussian ansatz} for the 
longitudinal part of the LFWF and $m_1$ and $m_2$ 
as constituent quark masses. In Ref.~\cite{Gutsche:2012ez} we followed  
Refs.~\cite{'tHooft:1974hx,Bergknoff:1976xr} and considered current 
quark masses. By an appropriate choice of the longitudinal wave 
function $f(x,m_1,m_2)$, we got consistency with QCD in both sectors 
of light and heavy quarks. In this way we generate the masses of light 
pseudoscalar mesons in agreement with the scheme resulting from 
explicit breaking of chiral symmetry --- in the leading order of 
the chiral expansion the masses of pseudoscalar mesons are linear 
in the current quark mass~\cite{Weinberg:1978kz}. 
We also guarantee that the pseudoscalar $\pi$, $K$, and $\eta$ meson masses 
$M_\pi^2, M_K^2, M_\eta^2$ satisfy the Gell-Mann-Oakes-Renner 
\eq 
M_\pi^2 = 2 \hat{m}  B 
\en 
and the Gell-Mann-Okubo 
\eq 
4 M_K^2 \,=\, M_\pi^2 \, + \, 3 M_\eta^2  
\en
relations, where $\hat{m} = (m_u + m_d)/2$ is the average mass 
of $u$ and $d$ quarks,  
$B  = | \la 0 | \bar u u | 0 \ra |/F_\pi^2$ is the quark condensate 
parameter, and $F_\pi$ is the leptonic decay constant. 
In the sector of heavy quarks we set agreement with both heavy quark effective 
theory and potential models of heavy quarkonia.  
In the heavy quark mass limit 
$m_Q \to \infty$ we obtaine the correct scaling of 
the leptonic decay constants for both heavy-light mesons 
$f_{Q\bar q} \sim 1/\sqrt{m_Q}$ and heavy quarkonia  
$f_{Q\bar Q} \sim \sqrt{m_Q}$ and 
$f_{c\bar b} \sim m_c/\sqrt{m_b}$ at $m_c \ll m_b$. 
In this limit we also generated the correct expansion of heavy meson masses 
\eq 
M_{Q\bar q} &=& m_Q + \bar\Lambda + {\cal O}(1/m_Q) \,, \nonumber\\
M_{Q\bar Q} &=& 2 m_Q + E + {\cal O}(1/m_Q) \,, 
\en 
where $\bar\Lambda$ is the approximate difference between the masses 
of the heavy-light meson and the heavy quark, $E$ is the binding energy 
in heavy quarkonia. The corresponding mass splittings, e.g. between vector 
and pseudoscalar states of heavy-light mesons, become  
\eq 
M_{Q\bar q}^V - M_{Q\bar q}^P \, \sim \, \frac{1}{m_Q} \,. 
\en  
We chose the longitudinal wave function in the form 
\eq\label{LLWF} 
f(x,m_1,m_2) = N \, x^{\alpha_1} \, (1-x)^{\alpha_2} \, 
\en 
where $N$ is the normalization constant fixed from
\eq
1 = \int\limits_0^1 dx \, f^2(x,m_1,m_2) 
\en
and $\alpha_1, \alpha_2$ are parameters that 
have been fixed in order to get consistency with QCD. 

Our master formula for the mass spectrum $M^2_{nJ}$ of quark-antiquark mesons 
in terms of the arbitrary longitudinal wave function $f(x,m_1,m_2)$ 
is given by the expression 
\eq\label{M2_mesons} 
M^2_{2q; nJ} = 4 \kappa^2 \biggl( n + \frac{L + J}{2} \biggr) 
+ \int\limits_0^1 dx
\biggl( \frac{m_1^2}{x} + \frac{m_2^2}{1-x} \biggr) f^2(x,m_1,m_2) \,.  
\en
Using our ansatz for the $f(x,m_1,m_2)$~(\ref{LLWF}) we get an analytic 
expression for the correction to the mass spectrum~\cite{Branz:2010ub}: 
\eq\label{M2_mesons_results} 
M^2_{2q; nJ} = 4 \kappa^2 \biggl( n + \frac{L + J}{2} \biggr) 
         + (1 + 2 \alpha_1 + 2 \alpha_2) \, 
\biggl( \frac{m_1^2}{2\alpha_1} 
      + \frac{m_2^2}{2\alpha_2} 
\biggr)\,. 
\en
In the case of the tetraquark states we have two possibilities: 
1) tetraquarks are compact bound states of two quarks and two antiquarks; 
2) tetraquarks are bound states of two mesons (hadronic molecules).

When tetraquarks have the hadronic molecular configuration the mass formula reads 
\eq\label{M4_our} 
(M_{4q; nJ}^{\rm HM})^2 &=& 
4 \kappa^2 \biggl( n + \frac{L + J}{2} + 1\biggr)
+\int\limits_0^1 dx
\biggl( \frac{M_1^2}{x} + \frac{M_2^2}{1-x} \biggr) f^2(x,M_1,M_2) 
\nonumber\\
&=& 4 \kappa^2 \biggl( n + \frac{L + J}{2} + 1\biggr)
         + (1 + 2 \alpha_1 + 2 \alpha_2) \, 
\biggl( \frac{M_1^2}{2\alpha_1} 
      + \frac{M_2^2}{2\alpha_2} 
\biggr)
\,,  
\en
where $M_1$ and $M_2$ are the masses of the quark-antiquark clusters forming the 
hadronic molecule. Note that $M_1$ and $M_2$ are a bit smaller than the 
physical masses of the corresponding mesons because of the separation of the  
contribution of the longitudinal and transverse wave function of tetraquarks. 
Somehow, one can call $M_1$ and $M_2$ bare meson masses, because the transverse 
part of the wave function gives an additional contribution due to the confinement 
potential.

We will now consider the four-quark structure of tetraquark states. 
Note that in Ref.~\cite{Brodsky:2016yod} the mass formula for 
the tetraquark states, in terms of quark degrees of freedom, was 
obtained using superconformal algebra as 
\eq\label{M4_Brodsky} 
M^2_{4q; nJ} = 4 \kappa^2 \biggl( n + L + \frac{S}{2} + 1 \biggr)
         + \frac{\kappa^4}{F(\kappa^2)} \, \frac{dF(\kappa^2)}{d\kappa^2} \,,
\en 
where 
\eq 
F(\kappa^2) = \int\limits_0^1 dx_1 \ldots \int\limits_0^1 dx_4 
\, \delta\biggl(\sum\limits_{i=1}^4 x_i - 1\biggr) 
\, \exp\biggl[- \sum\limits_{i=1}^4 \frac{m_i^2}{x_i\kappa^2} \biggr] \,. 
\en 
The difference between the two formulas~(\ref{M4_our}) 
and~(\ref{M4_Brodsky}) in the zero mass limit for the constituents is due 
to the term which spontaneously breaks superconformal symmetry 
\eq 
\Delta M^2_{4q; nJ} = 2 \kappa^2 \, (J-L-S) \,, 
\en 
which is zero for tetraquark systems with $J = L + S$. 

Notice that our formula gives, in the zero mass limit, an natural  
explanation of the tetraquark state $a(1420)$ discovered 
by the COMPASS Collaboration~\cite{Adolph:2015pws}. 
In our approach, for $n=0$, $J = L = 1$, we get
\eq 
M^2_{a(1420)} = 8 \kappa^2 
\en   
or $M_{a(1420)}  = 2 \kappa \sqrt{2}$ resulting in  
$M_{a(1420)}  = \sqrt{2}$ GeV $\simeq 1.414$ GeV at $\kappa = 0.5$ GeV, 
which agrees perfectly with the experimental result of  
$M_{a(1420)}  = 1.414^{+0.015}_{-0.013}$ GeV. 
The analysis of Ref.~\cite{Wang:2014bua} within QCD sum rules  
disfavors assigning the $a_1(1420)$ to an axial-vector tetraquark state 
and proposes that it is a mixed state of the $a_1(1260)$ meson 
and the tetraquark state with the configuration 
$[su]_{S=1} [\bar s \bar d]_{S=0} + [su]_{S=0} [\bar s \bar d]_{S=1}$. 
On the other hand, 
the QCD sum rules analysis performed in Ref.~\cite{Chen:2015fwa} 
confirmed the existence of  $a_1(1420)$ as a tetraquark state. 
Questions related to the nature of the $a_1(1420)$ meson 
have also been addressed 
in other papers. In particular, it has been proposed that this state 
is a consequence of rescattering effects, and in fact 
in Ref.~\cite{Ketzer:2015tqa} the $a_1(1420)$ was interpreted as 
a dynamical effect due to a singularity (branch point) in the 
triangle diagram formed by the processes 
$a_1(1260) \to K^\ast \bar K$, $K^\ast \to K \pi$ 
and $K \bar K \to f_0(980)$. 
In Ref.~\cite{Basdevant:2015wma} it was shown that 
a single $I=1$ spin-parity $J^{PC}=1^{++}$ $a_1$ resonance can manifest itself 
as two separated mass peaks. One decays into an $S$-wave $\rho\pi$ system 
and the second decays into a $P$-wave $f_0(980)\pi$ system, with a rapid 
increase of the phase difference between their amplitudes, arising mainly from 
the structure of the diffractive production process. 
In Ref.~\cite{Liu:2015taa} it was claimed that resonances such as the $a_1(1420)$ 
could be produced due to the so-called ``anomalous triangle singularity'', 
if it is located in a specific kinematical region. 
In Ref.~\cite{Aceti:2016yeb} $a_1(1420)$ was considered as 
peak in the $a_1(1260) \to \pi f_0(980)$ decay mode. 
In Ref.~\cite{Wang:2015cis} it was proposed to test the possible 
rescattering nature of the $a_1(1420)$ in heavy meson decays. 

Our approach gives also a lower 
limit for the masses of the tetraquarks with 
$n = J = L = S = 0$ 
\eq 
M_{4q} \ge 2 \kappa = 1 \, {\rm GeV}
\en 
for $\kappa = 0.5$ GeV. It means that the lightest possible 
tetraquark is composed of two diquarks forming a state 
with spin parity $J^P = 0^+$, and then a possible candidate is 
the $f_0(980)$ consistent with the
conclusion of Ref.~\cite{Brodsky:2016yod}.  
Note that in the massless limit we consider the same universal 
dilaton parameter $\kappa$ for the constituents of all states forming 
tetraquarks. When the masses of the tetraquark constituents 
are taken into account, we assume that the numerical value of 
$\kappa$ for tetraquarks could change from its original value of $0.5$ GeV. 

Next we include quark mass effects, following the approach presented 
in our paper on applications of holographic QCD 
to mesons~\cite{Gutsche:2012ez}. We proposed the following form 
for the longitudinal wave function containing quark masses 
\eq\label{f4q} 
f(x_1,\ldots,x_4,m_1,\ldots,m_4) = 
N \, x_1^{\alpha_1} \ldots x_4^{\alpha_4}, 
\en
where $N$ is a normalization constant fixed from the condition 
\eq 
1 = \int\limits_0^1dx_1 \cdots \int\limits_0^1dx_4 \, 
\delta\Big(\sum\limits_{i=1}^4 x_i - 1\Big) \, 
f^2(x_1,\ldots,x_4,m_1,\ldots,m_4) \,. 
\en 
The contribution of the longitudinal wave function~(\ref{f4q}) 
to the mass spectrum is
\eq 
\Delta M^2_{4q; nJ} &=& \int\limits_0^1dx_1 \cdots \int\limits_0^1dx_4 \, 
\delta\Big(\sum\limits_{i=1}^4 x_i - 1\Big) \, 
f^2(x_1,\cdots,x_4,m_1,\cdots,m_4) \sum\limits_{i=1}^4 \, \frac{m_i^2}{x_i} 
\nonumber\\
&=& \Big(3 + 2 \sum\limits_{i=1}^4 \alpha_i\Big) \, 
\sum\limits_{i=1}^4 \frac{m_i^2}{2 \alpha_i} \,. 
\en 
It should be stressed, as in case of mesons, that 
by an appropriate choice of 
the $\alpha_i$ parameters we can guarantee the correct 
behavior of the tetraquark spectrum in both the light and heavy quark sectors. 
In particular, tetraquarks composed of light nonstrange $(u,d)$ quarks 
receive the following quark mass correction 
\eq 
\Delta M^2_{4q; nJ}([q\bar q]^2) = 2 \Big( \frac{3}{\alpha_q} + 8 \Big) 
\, \hat{m}^2 \,, 
\en 
where $\hat{m} = m_u = m_d$ in the isospin limit. 
The parameter $\alpha_q = 3\hat{m}/(2B)$ is fixed from the condition 
\eq
\Delta M^2_{4q; nJ}([q\bar q]^2) = 
4 \hat{m} B + {\cal O}(\hat{m}^2) \simeq 2 M_\pi^2 \,. 
\en 
By analogy, for light tetraquarks containing single, two and three 
strange quarks/antiquarks 
using $\alpha_q = 3\hat{m}/(2B)$ and $\alpha_s = 3m_s/(2B)$ we get 
\eq 
\Delta M^2_{4q; nJ}([s\bar q][q\bar q]) &=& 
\Delta M^2_{4q; nJ}([q\bar s][q\bar q]) \ = \ 
(3 + 6\alpha_q + 2\alpha_s) \Big( \frac{3\hat{m}^2}{2\alpha_q} 
+ \frac{m^2_s}{2\alpha_s}\Big)\nonumber\\ 
&=& 
B (3\hat{m} + m_s) + {\cal O}(\hat{m}^2,m_s^2,\hat{m}m_s)  
\simeq M_\pi^2 + M_K^2\,, 
\en
\eq 
\Delta M^2_{4q; nJ}([q\bar s]^2) &=& 
\Delta M^2_{4q; nJ}([s\bar q]^2) \ = \  
\Delta M^2_{4q; nJ}([s\bar q][q\bar s]) \ = \
(3 + 4\alpha_q + 4\alpha_s) \Big( \frac{\hat{m}^2}{\alpha_q} 
+ \frac{m^2_s}{\alpha_s}\Big)\nonumber\\
&=& 2 B (\hat{m} + m_s) 
+ {\cal O}(\hat{m}^2,m_s^2,\hat{m}m_s)   
\simeq 2 M_K^2\,, 
\en   
\eq 
\Delta M^2_{4q; nJ}([q\bar s][s\bar s]) &=& 
\Delta M^2_{4q; nJ}([s\bar q][s\bar s]) \ = \ 
(3 + 2\alpha_q + 6\alpha_s) \Big( \frac{\hat{m}^2}{2\alpha_q} 
+ \frac{3m^2_s}{2\alpha_s}\Big) \nonumber\\
&=& B (\hat{m} + 3m_s) + {\cal O}(\hat{m}^2,m_s^2,\hat{m}m_s)  
\simeq 3 M_K^2 - M_\pi^2\,, 
\en    
and 
\eq 
\Delta M^2_{4q; nJ}([s\bar s]^2) &=& 
2 \Big( \frac{3}{\alpha_s} + 8 \Big) \, m_s^2 \nonumber\\
&=& 4 m_s B 
+ {\cal O}(m_s^2) \simeq 4 M_K^2 - 2 M_\pi^2 \,.  
\en   
We will now derive quark mass corrections for tetraquarks containing single, 
two, tree and four heavy quarks/anti\-quarks $Q=b,c$ 
(in the following index $q$ denotes all 
light quarks $u,d,s$, and for simplicity we consider heavy and 
light quarks each of the same flavor): 
\eq 
\Delta M^2_{4q; nJ}([Q\bar q] [q\bar q]) &=&  
\Delta M^2_{4q; nJ}([q\bar Q] [q\bar q]) \ = \  
(3 + 2\alpha_Q + 6\alpha_q) 
\Big( \frac{m_Q^2}{2\alpha_Q} + \frac{3m_q^2}{2\alpha_q}\Big)  
\,, \nonumber\\ 
\Delta M^2_{4q; nJ}([Q\bar Q] [q\bar q]) &=& 
\Delta M^2_{4q; nJ}([Q\bar q] [q\bar Q]) \ = \ 
(3 + 4\alpha_Q + 4\alpha_q) \Big( \frac{m_Q^2}{\alpha_Q} 
+ \frac{m_q^2}{\alpha_q}\Big)  
\,, \nonumber\\ 
\Delta M^2_{4q; nJ}([Q\bar q] [Q\bar Q]) &=&  
\Delta M^2_{4q; nJ}([q\bar Q] [Q\bar Q]) \ = \  
(3 + 6\alpha_Q + 2\alpha_q) 
\Big( \frac{3m_Q^2}{2\alpha_Q} + \frac{m_q^2}{2\alpha_q}\Big)  
\,, \nonumber\\ 
\Delta M^2_{4q; nJ}([Q\bar Q]^2) &=& 
16 m_Q^2 \biggl(1 + \frac{3}{8\alpha_Q}\biggr) \,. 
\en
To get the correct scaling of heavy tetraquarks for 
 $m_Q \to \infty$ with 
\eq 
\Delta M^2_{4q; nJ}([Q\bar q] [q\bar q]) &= & 
(m_Q + \bar\Lambda_{3q} + {\cal O}(1/m_Q))^2\,, 
\nonumber\\
\Delta M^2_{4q; nJ}([Q\bar Q] [q\bar q]) &=& 
\Delta M^2_{4q; nJ}([Q\bar q] [q\bar Q]) \ = \ 
(2m_Q + \bar\Lambda_{2q} + {\cal O}(1/m_Q))^2\,, 
\nonumber\\
\Delta M^2_{4q; nJ}([Q\bar Q] [Q\bar q]) &=&  
\Delta M^2_{4q; nJ}([q\bar Q] [Q\bar Q]) \ = \ 
(3m_Q + \bar\Lambda_q + {\cal O}(1/m_Q))^2\,, 
\nonumber\\
\Delta M^2_{4q; nJ}([Q\bar Q] [Q\bar Q]) &=&  
(4m_Q + E + {\cal O}(1/m_Q))^2  \,,
\en 
where $\bar\Lambda_{nq}$ is the approximate difference between the mass
of the heavy-light tetraquark and the sum of the masses of its $(4-n)$ 
constituent heavy quarks, we fix the $\alpha_Q$ parameters as follows: 
$\alpha_Q = \alpha$ (independent on the heavy quark flavor) 
in the case of heavy-light tetraquarks and 
\eq 
\alpha_Q = \frac{3}{4} \, \frac{m_Q}{E} \biggl( 1 - \frac{E}{8m_Q} \biggr)
\en 
in the case of tetraquarks composed only of heavy quarks. The parameter for 
light quarks $\alpha_q$ occuring in the case of heavy-light tetraquarks 
is fixed as 
\eq 
\alpha_q \equiv \alpha_q^{(n)} 
= \frac{2 \alpha}{n} \, \frac{\bar\Lambda_{nq}}{m_Q} \, 
\biggl( 1 + \frac{1}{2 (4-n)}\frac{\bar\Lambda_{nq}}{m_Q} \biggr) 
- \frac{3}{2n}
\en 
where $n$ is the number of light quark/antiquarks in a tetraquark. 
In particular, 
\eq 
\alpha_q^{(3)} 
= \frac{2 \alpha}{3} \, \frac{\bar\Lambda_{3q}}{m_Q} \, 
\biggl( 1 + \frac{\bar\Lambda_{3q}}{2 m_Q} \biggr) 
- \frac{1}{2}
\en 
in the case of $n=3$, 
\eq
\alpha_q^{(2)} 
= \alpha \, \frac{\bar\Lambda_{2q}}{m_Q} \, 
\biggl( 1 + \frac{\bar\Lambda_{2q}}{4 m_Q} \biggr) 
- \frac{3}{4} 
\en 
in the case of $n=2$, 
\eq 
\alpha_q^{(1)} 
= 2 \alpha \, \frac{\bar\Lambda_{q}}{m_Q} \, 
\biggl( 1 + \frac{\bar\Lambda_{q}}{6m_Q} \biggr) - \frac{3}{2}
\en  
in the case of $n=1$. 

In the case of tetraquarks with a hadronic molecular configuration
it is worthwhile to test the prediction of Ref.~\cite{Bondar:2011ev} 
about the structure of the $Z_b(10610) = 
(B \bar B^\ast + {\rm h.c.})$ and $Z_b(10650) = (B^\ast \bar B^\ast)$ 
resonances, which are supposed to be $J^P = 1^+$ states. 
In our approach their masses are given by 
\eq 
M_{Z_b(10610)}^2 &=& 8 \kappa^2 + (1 + 2 \alpha_{P_{q\bar b}} 
+ 2 \alpha_{V_{q\bar b}}) 
\, \biggl( \frac{M_{P_{q\bar b}}^2}{2 \alpha_{P_{q\bar b}}} 
         + \frac{M_{V_{q\bar b}}^2}{2 \alpha_{V_{q\bar b}}}  \biggr)  \,, 
\nonumber\\
M_{Z_b(10650)}^2 &=& 8 \kappa^2 + (1 + 4 \alpha_{V_{q\bar b}}) 
\, \frac{M_{V_{q\bar b}}^2}{\alpha_{V_{q\bar b}}}  \,.  
\en 
Fixing $\alpha_{P_{q\bar b}}$ and $\alpha_{V_{q\bar b}}$ as 
\eq\label{alphaB}  
\alpha_i = \frac{1}{4} \biggl(\frac{M_i}{\Delta_B} - \frac{1}{4}\biggr) \,, \quad 
\Delta_B = M_{V_{q\bar b}} - M_{P_{q\bar b}} \equiv M_{B^\ast} - M_B \,
\en 
results in the expressions 
\eq 
M_{Z_b(10610)}^2 &=& 8 \kappa^2 \, + \, 
 (M_{V_{q\bar b}}  + M_{P_{q\bar b}} + \Delta_B)^2 \, 
+ {\cal O}(\Delta_B^3) = 8 \kappa^2 \, + \, 
4 \biggl(\bar M_B + \frac{\Delta_B}{2}\biggr)^2 \, + \, {\cal O}(\Delta_B^3) 
\,,\nonumber\\
M_{Z_b(10650)}^2 &=& 8 \kappa^2 + 
\, (2 M_{V_{q\bar b}} + \Delta_B)^2 \, + {\cal O}(\Delta_B^3) 
= 8 \kappa^2 \, + \, 
4 (\bar M_B +\Delta_B)^2 
\, + \, {\cal O}(\Delta_B^3) \,, 
\en
where $\bar M_B = \frac{M_{V_{q\bar b}} + M_{P_{q\bar b}}}{2}$. 
For the mass splitting $M_{Z_b(10650)}^2 - M_{Z_b(10610)}^2$ 
we get 
\eq 
M_{Z_b(10650)}^2 - M_{Z_b(10610)}^2 
= 4 \Delta_B \bar M_B + {\cal O}(\Delta_B^2) 
\en 
or 
\eq 
\Delta = M_{Z_b(10650)} - M_{Z_b(10610)} = \Delta_B + {\cal O}(\Delta_B^2)
\en 
in agreement with Ref.~\cite{Bondar:2011ev}. By analogy we get a similar 
conclusion for the related charmed  
tetraquark states $Z_c(3900) = (D \bar D^\ast + {\rm h.c.})$ and
$Z_c(4020) = (D^\ast \bar D^\ast)$. 
  
Another interesting state is the $X(3872)$. In our approach we consider 
this state in a mixed configuration of the molecular 
$(D^0\bar D^{\ast 0} + \bar D^0  D^{\ast 0})/\sqrt{2}$ 
and the charmonia $c\bar c$ states following 
Refs.~\cite{Swanson:2003tb,Dong:2008gb} with 
\eq\label{X3872} 
|X(3872)\ra = \frac{\cos\theta}{\sqrt{2}} \, 
|D^0 \bar D^{\ast 0} + \bar D^0  D^{\ast 0}\ra \, 
+ \, \sin\theta \, \bar cc \,.
\en   
Using Eq.~(\ref{X3872}) 
and 
\eq 
\alpha_c = \frac{1}{4} \biggl( \frac{m_c}{E} - \frac{1}{4}\biggr) \,, \quad 
\alpha_i = \frac{1}{4} \biggl( \frac{M_i}{\Delta_D} - \frac{1}{4}\biggr)\,, 
\quad i = V_{c\bar q}, P_{c\bar q} 
\en 
we get the following mass formula for 
the $X(3872)$ state 
\eq 
M_{X(3872)}^2 &=& \cos^2\theta \biggl[ 
8 \kappa^2 + 
(1 + 2 \alpha_{P_{c\bar q}}
+ 2 \alpha_{V_{c\bar q}})
\, \biggl( \frac{M_{P_{c\bar q}}^2}{2 \alpha_{P_{c\bar q}}}
         + \frac{M_{V_{c\bar q}}^2}{2 \alpha_{V_{c\bar q}}}\biggr) 
\biggr] \, + \, 
\sin^2\theta \biggl[ 4 \kappa^2 + 4 m_c^2 \biggl(1 + \frac{1}{4\alpha_c}\biggr) 
\biggr] \nonumber\\
&=& 4 \kappa^2 (1 + \cos^2\theta)  
\, + \, 4 \biggl(\bar M_D + \frac{\Delta_D}{2}\biggr)^2 \, \cos^2\theta \, 
\, + \, (2 m_c + E)^2 \, \sin^2\theta 
\, + \, {\cal O}(\Delta_D^3, E^3)\,,
\en  
where by analogy with the bottom sector we have 
\eq 
\bar M_D = \frac{M_{V_{c\bar q}} + M_{P_{c\bar q}}}{2}\,, \quad 
\Delta_D = M_{V_{c\bar q}} - M_{P_{c\bar q}} \equiv M_{D^\ast} - M_D \,. 
\en 
Note that the hadronic molecular contribution to the $X(3872)$ state 
is the same expression as for the $Z_c(3900)$ and $Z_c(4020)$ states. 
The values $m_c = 1.275$ GeV and $E = 0.795$ GeV are 
taken from Ref.~\cite{Gutsche:2012ez}.  
We find that for $|\theta| = 11^0$ we could reproduce the experimental 
value of the $X(3872)$ mass of $M_{X(3872)} = 3871.69 \pm 0.17$ MeV. 

Next we look at the possible $J^P = 0^+$ $Y$--states as hadronic molecules 
\eq 
|Y(3940)\ra &=& \frac{1}{\sqrt{2}} \, 
| D^{\star +} D^{\star -} + D^{\star 0} \bar D^{\star 0} \ra \,, 
\nonumber\\
|Y(4140)\ra &=&  | D_s^{\star +} D_s^{\star -}  \ra
\en 
studied by us before using a phenomenological 
Lagrangian approach in Ref.~\cite{Branz:2009yt}. 
Holographic QCD gives the following results for the masses of 
$Y(3940)$ and $Y(4140)$ states 
\eq 
M_{Y(3940)}^2 &=& 
4 \kappa^2 + 4 M_{V_{c\bar q}}^2 \biggl(1 + \frac{1}{4 \alpha_{Y(3940)}}\biggr) 
\nonumber\\
&=& 4 \kappa^2 + 4 (M_{V_{c\bar q}} + \Delta_Y)^2 \,, 
\nonumber\\
M_{Y(4140)}^2 &=& 
4 \kappa^2 + 4 M_{V_{c\bar s}}^2 \biggl(1 + \frac{1}{4 \alpha_{Y(4140)}}\biggr) 
\nonumber\\
&=& 4 \kappa^2 + 4 (M_{V_{c\bar s}} + \Delta_Y)^2 \,, 
\en
where 
\eq 
\alpha_{Y(3940)} =  \frac{1}{8} \, \biggl( \frac{M_{V_{c\bar q}}}{\Delta_Y} 
- \frac{1}{2} \biggr)\,, 
\quad 
\alpha_{Y(4140)} =  \frac{1}{8} \, \biggl( \frac{M_{V_{c\bar s}}}{\Delta_Y} 
- \frac{1}{2} \biggr)\,, 
\en 
and $\Delta_Y$ is phenomenological parameter of order of ${\cal O}(M_i^0)$. 
With $\Delta_Y = 96$ MeV we can reproduce the mass data for these states. 

Finally, we consider the $Z(4430)^+$ state 
in the hadronic molecular picture, which has the structure~\cite{Branz:2010sh} 
\eq
|Z(4430)^+\ra = \frac{1}{\sqrt{2}} \, | D_1^{+} \bar D^{\ast 0} 
+ D^{\star +} \bar D_1^{0} \ra \,.
\en 
Its mass squared neglecting isospin breaking corrections 
is given by 
\eq 
M_{Z(4430)}^2 = 8 \kappa^2 + 
\biggl( 1 + 2 \alpha_{A_{c\bar q}} + 2 \alpha_{V_{c\bar q}}\biggr) 
\biggl(    \frac{M_{A_{c\bar q}}^2}{2 \alpha_{A_{c\bar q}}}
         + \frac{M_{V_{c\bar q}}^2}{2 \alpha_{V_{c\bar q}}}  \biggr)\,. 
\en 
Taking 
\eq 
\alpha_i = \frac{1}{2} \biggl(\frac{M_i}{\Delta_Z} - \frac{1}{8} \biggr) \,, \quad 
\Delta_Z = M_{A_{c\bar q}} - M_{V_{c\bar q}} \equiv M_{D_1} - M_{D^\ast} 
\en 
we get 
\eq 
M_{Z(4430)}^2 &=& 8 \kappa^2 \, + \, 
\biggl(M_{A_{c\bar q}}  + M_{V_{c\bar q}} + \frac{\Delta_Z}{2}\biggr)^2 \, 
+ {\cal O}(\Delta_Z^3) \,. 
\en 
Our predictions for the masses for selected light and heavy tetraquarks 
are shown in Tables~1-4. For some states, suspected as tetraquarks, 
we include a comparison with data~\cite{Olive:2016xmw}, while
for other states we make mass predictions.  
In future work we plan to make a more detailed analysis. 

For completeness we also consider 
two candidates on tetraquarks with color diquark-color diquark 
configuration --- $X(3755)$ and $X(3915)$ states,  
with $\bar\Lambda_{2q} = 1.07$ GeV we reproduce the data for their masses. 
Also we suppose that $X(3915)$ is the radial excitation of the $X(3755)$ state. 
 
In the numerical analysis we use the same set of quark masses as 
for the conventional quark-antiquark mesons: 
\eq\label{const_qm}
& &m_u = m_d = \hat{m} = 7 \ {\rm MeV}\,, \nonumber\\
& &m_s = 24 \hat{m} = 168 \ {\rm MeV}\,, \nonumber\\
& &m_c = 1.275 \ {\rm GeV}\,, \hspace*{.5cm}
   m_b = 4.18  \ {\rm GeV} \,. 
\en 
We use the universal
parameters $\kappa = 351/435$ MeV for light and $\kappa = 500$ MeV 
for heavy-light tetraquarks. 
We suppose that the inclusion of finite mass effects of the constituents 
in the light tetraquarks shifts the parameter $\kappa$ to lower values 
than 500 MeV. However, in the massless limit for the tetraquark constituents, 
the parameter $\kappa$ is universal and the same 
for all tetraquark states independent on the light/heavy quark/diquark/meson 
content. The masses of the quark-antiquark clusters for pseudoscalar states 
are taken as 
\eq 
M_{P_{c \bar q}} = 1.669 \ {\rm GeV} \,, \quad 
M_{P_{c \bar s}} = 1.772 \ {\rm GeV} \,, \quad 
M_{P_{u \bar b}} = 5.211 \ {\rm GeV} \,.
\en 
As we stressed before, the masses of other spin-parity states are calculated 
according to the mass splittings 
\eq 
& &M_{V_{c \bar q}} = M_{P_{c \bar q}} + M_{D^\ast} - M_D\,,      \quad 
   M_{V_{c \bar s}} = M_{P_{c \bar s}} + M_{D^\ast_s} - M_{D_s}\,,\nonumber\\
& &M_{V_{q \bar b}} = M_{P_{q \bar b}} + M_{B^\ast} - M_B\,,
\en 
and etc. 

\begin{acknowledgments} 

The authors thank Stan Brodsky for useful discussions. 
This work was funded
by the German Bundesministerium f\"ur Bildung und Forschung (BMBF)
under Project 05P2015 - ALICE at High Rate (BMBF-FSP 202):
``Jet- and fragmentation processes at ALICE and the parton structure 
of nuclei and structure of heavy hadrons'';
by CONICYT (Chile) PIA/Basal FB0821, 
by CONICYT (Chile) Research Projects No. 80140097,
and under Grants No. 7912010025, 1140390;
by Tomsk State University Competitiveness
Improvement Program and the Russian Federation program ``Nauka''
(Contract No. 0.1764.GZB.2017).

\end{acknowledgments}

\newpage 

\begin{widetext}

\begin{table}
\caption{Masses of possible light tetraquarks (in MeV).  \label{tab:1}}
\begin{tabular}{|c|c|c|c|c|c|c|}
\hline
Tetraquark & Quark Content 
& Quantum Numbers & $\kappa$ & Mass & Mass &
Data~\cite{Olive:2016xmw} \\
           &   & $(J^P,n,L,S)$   &  (MeV)  &  (Zero Quark Masses)  &   (Finite Quark Masses)  &  \\
\hline 
$f_0(980)$   & $[q \bar s] [s \bar q]$ & ($0^+$,0,0,0) & 351      & 702      & 990     & 990  $\pm$ 20   \\
$f_1(1215)$  & $[q \bar s] [s \bar q]$ & ($1^+$,0,1,0) & 351      & 993      & 1214    &  \\
$f_2(1400)$  & $[q \bar s] [s \bar q]$ & ($2^+$,0,2,0) & 351      & 1216     & 1402    &  \\
$f_3(1570)$  & $[q \bar s] [s \bar q]$ & ($3^+$,0,3,0) & 351      & 1404     & 1568    &  \\
$a_1(1420)$  & $[q \bar s] [s \bar q]$ & ($1^+$,0,1,1) & 435      & 1230     & 1414    & 
$1414^{+15}_{-13}$~\cite{Adolph:2015pws}\\
\hline
\end{tabular}

\vspace*{.5cm}

\caption{Masses of heavy tetraquarks (in MeV). \label{tab:2}}
\begin{tabular}{|c|c|c|c|c|c|}
\hline
Tetraquark & Quark Content 
& Quantum Numbers & $\kappa$ & Mass & 
Data~\cite{Olive:2016xmw} \\
           &   & $(J^P,n,L,S)$   &  (MeV)  &  & \\ 
\hline 
$X(3755)$& $ [c \bar q] [q \bar c]$  & ($0^+$,0,0,0) 
& 500   & 3756            & \\
$X(3915)$& $ [c \bar q] [q \bar c]$  & ($0^+$,1,0,0)
& 500   & 3886            & 3918.4  $\pm$  1.9 \\
$Y(3940)$& $ [c \bar q] [q \bar c]$  & ($0^+$,0,0,0)
& 500   & 3940            & 3943.0  $\pm$  11 $\pm$ 13 \\
$Y(4140)$& $ [c \bar s] [s \bar c]$  & ($0^+$,0,0,0)
& 500   & 4146            & 4143.0  $\pm$  2.9 $\pm$ 1.2 \\
$X(3872)$& $ \cos\theta [c \bar d] [d \bar c] + 
\sin\theta c \bar c$  & ($1^+$,0,1,1)
& 500   & 3872            & 3871.69  $\pm$  0.17 \\
$Z_c(3900)^+$& $[c \bar d] [u \bar c]$ & ($1^+$,0,1,1) 
& 500   & 3886            & 3886.6  $\pm$  2.4 \\
$Z_c(4020)^+$& $[c \bar d] [u \bar c]$  & ($1^+$,0,1,1) 
& 500   & 4017            & 4024.1  $\pm$  1.9 \\
$Z(4430)^+$& $[c \bar d] [u \bar c]$  & ($1^+$,0,1,1) 
& 500   & 4468            & $4478^{+15}_{-18}$ \\
$Z_b(10610)^+$& $ [b \bar d] [u \bar b]$  & ($1^+$,0,1,1)
& 500   & 10607          & 10607.2  $\pm$  2.0 \\
$Z_b'(10650)^+$& $ [b \bar d] [u \bar b]$  & ($1^+$,0,1,1)
& 500   & 10652           & 10652.2  $\pm$  1.5 \\
\hline
\end{tabular}
\end{table}

\end{widetext}

\end{document}